\documentclass[letterpaper]{article} 
\usepackage{aaai24}  
\usepackage{times}  
\usepackage{helvet}  
\usepackage{courier}  
\usepackage[hyphens]{url}  
\usepackage{graphicx} 
\usepackage{natbib}  
\usepackage{caption} 
\usepackage{algorithm}
\usepackage{algorithmic}

\usepackage{amsmath,amsfonts}
\usepackage{pifont}
\usepackage[caption=false,font=normalsize,labelfont=sf,textfont=sf]{subfig}
\usepackage{multirow}
\usepackage{xspace}

\newcommand{\ie}{\textit{i.e.}\xspace}
\newcommand{\eg}{\textit{e.g.}\xspace}

\usepackage{color}

\usepackage{newfloat}
\usepackage{listings}
\DeclareCaptionStyle{ruled}{labelfont=normalfont,labelsep=colon,strut=off} 
\floatstyle{ruled}
\newfloat{listing}{tb}{lst}{}
\floatname{listing}{Listing}
\title{Towards Model Extraction Attacks in GAN-Based Image Translation \\ via Domain Shift Mitigation}

\author {
    Di Mi\textsuperscript{\rm 1},
    Yanjun Zhang\textsuperscript{\rm 2},
    Leo Yu Zhang\textsuperscript{\rm 3},\\
    Shengshan Hu\textsuperscript{\rm 4},
    Qi Zhong\textsuperscript{\rm 5},
    Haizhuan Yuan\textsuperscript{\rm 1},
    Shirui Pan\textsuperscript{\rm 3}
}
\affiliations {
    \textsuperscript{\rm 1}Xiangtan University\\
    \textsuperscript{\rm 2}University of Technology Sydney\\
    \textsuperscript{\rm 3}Griffith University\\
    \textsuperscript{\rm 4}Huazhong University of Science and Technology\\
    \textsuperscript{\rm 5}City University of Macau\\
    midi\_xtu@163.com, 
    yanjun.zhang@uts.edu.au, 
    leo.zhang@griffith.edu.au, \\
    hushengshan@hust.edu.cn,
    qizhong@cityu.edu.mo,
    yhz@xtu.edu.cn,
    s.pan@griffith.edu.au
   
}

\usepackage{bibentry}

\begin{document}

\maketitle

\begin{abstract}
Model extraction attacks (MEAs) enable an attacker to replicate the functionality of a victim deep neural network (DNN) model by only querying its API service remotely, posing a severe threat to the security and integrity of pay-per-query DNN-based services. Although the majority of current research on MEAs  has primarily concentrated on neural classifiers, there is a growing prevalence of image-to-image translation (I2IT) tasks in our everyday activities. However, techniques developed for MEA of DNN classifiers cannot be directly transferred to the case of I2IT, rendering the vulnerability of I2IT models to MEA attacks often underestimated. This paper unveils the threat of MEA in I2IT tasks from a new perspective. Diverging from the traditional approach of bridging the distribution gap between attacker queries and victim training samples, we opt to mitigate the effect caused by the different distributions, known as the domain shift. This is achieved by introducing a new regularization term that penalizes high-frequency noise, and seeking a flatter minimum to avoid overfitting to the shifted distribution. Extensive experiments on different image translation tasks, including image super-resolution and style transfer, are performed on different backbone victim models, and the new design consistently outperforms the baseline by a large margin across all metrics. A few real-life I2IT APIs are also verified to be extremely vulnerable to our attack, emphasizing the need for enhanced defenses and potentially revised API publishing policies.
\end{abstract}

\section{Introduction}
\label{sec:Introduction}

Deep neural networks (DNNs) have exhibited remarkable success in diverse domains, driven by substantial investments in data processing, computational power, {and expertise knowledge~\cite{pouyanfar2018survey}}. This success has been capitalized through the introduction of pay-per-query API services\footnotemark[4]. However, recent research has uncovered a notable vulnerability: model extraction attacks (MEAs). These attacks empower an adversary to replicate the remote DNN's functionality by crafting {surrogate models~\cite{mea_APIs_equation}}. Consequently, this allows unauthorized access to the service, enabling the adversary to launch adversarial attacks or privacy attacks on the service provider~\cite{xyrw_black-box_attack1,zhaoximembership22,mengyaoloden23,xyrw_membership_inference}. Such vulnerabilities not only undermine the security of the entire model supply chain but also pose a critical challenge to the integrity of DNN-based services.

Despite the apparent simplicity of this attack process, a significant challenge persists for the adversary: the queried samples are unlikely to perfectly match the secret training dataset utilized for training the victim model, then how can the MEA be executed in a manner that ensures both effectiveness and efficiency, given this discrepancy between queried and training samples?

To close the distribution gap between query and training samples, numerous works have been proposed for various classification tasks, including image  processing~\cite{mea_knockoff,mea_activethief,
mea_ES_attack,mea_confidence}, NLP processing~\cite{mea_nlp_first,
mea_nlp_better}, and graph data processing~\cite{mea_gnn1,mea_gnn3}. For instance, in image classification, \cite{mea_knockoff} suggested using reinforcement learning, and \cite{mea_activethief} suggested using active learning as a means to identify better query samples. Alternatively, \cite{mea_ES_attack,mea_confidence} advocated aligning the distribution of the secret training dataset with Generative Adversarial Networks (GANs). In NLP tasks, \cite{mea_nlp_better} suggested targeting an ensemble of victim APIs simultaneously, all providing the same service, to make them implicitly vote for good query samples. 

Concurrent with classification tasks, image-to-image translation~(I2IT) tasks, such as image super-resolution or restoration, constitute another significant application domain of DNNs. As a testament to this prevalence, Table~\ref{tab:price} makes a comparison of the pay-per-use price between classification and I2IT tasks on different platforms\footnotemark[1]$^{,}$\footnotemark[2].
Nonetheless, the MEA vulnerabilities in I2IT are rarely explored, making their risk largely underestimated. 

\footnotetext[1]{Providing image translation service: https://cloud.baidu.com,  https://imglarger.com, https://vanceai.com.}
\footnotetext[2]{Providing classification service: https://cloud.baidu.com, https://aws.amazon.com, https://cloud.google.com.}

We attribute this to the natural difference between studying MEA on classification models and I2IT models. In particular, extracting a classification model corresponds to effectively identifying its decision boundary~{\cite{mea_activethief}}. Hence, any alteration in the label or confidence scores of a queried sample (results yielded by the victim API)  implies the sample is crossing the decision boundary or changing its proximity to the boundary, which can be exploited directly by the adversary to select better queries.
In contrast, I2IT models take images as inputs and produce images as their outputs. It remains unknown what kind of information regarding the victim model the output images carry, and which output image carries more. 

Considering this discrepancy, this paper introduces an innovative approach to initiate MEA on I2IT models. The core idea is to directly mitigate the domain shift problem when training the surrogate. To achieve this, we design two complemented components: one regulates the behavior of the surrogate to suppress noisy components in translation outcomes while reducing model complexity, and the other pursues a flatter optimum to avoid overfitting to the shifted distribution. To our best knowledge, this is the first time that domain shift mitigation techniques is investigated in the context of MEAs. The contribution of this work is twofold: 
\begin{itemize}
    \item 
    Besides the traditional wisdom of closing the distribution discrepancy in MEA, we, for the first time, highlight that mitigating the domain shift constitutes another angle for launching MEA attacks. This approach proves especially advantageous in scenarios where how to select better queries is not clear. This fresh angle on MEA attacks is of independent research interest. 
    \item 
    We apply concrete domain shift mitigation strategies (\ie, wavelet regularization and sharpness-aware minimization) to extract GAN-based models in I2IT tasks. Extensive experimental results in controlled laboratory conditions and real-world scenarios corroborate that MEA is a real threat to image translation systems. 
\end{itemize}

\begin{table}[t]
\centering
\fontsize{9pt}{11pt}\selectfont
\begin{tabular}{|c|c|c|}
\hline
     \multicolumn{3}{|c|}{Image translation}\\
\hline
     Baidu AI Cloud & Imglarger & VanceAI \\
\hline
     \textdollar{0.0064-0.0614} & \textdollar{0.09} & \textdollar{0.035} \\
\hline    
     \multicolumn{3}{|c|}{Classification}\\
\hline
     Amazon & Google Cloud  & Baidu AI Cloud\\
\hline
     \textdollar{0.0008-0.001} & \textdollar{0.001-0.0015} & \textdollar{0.00029-0.00041}\\
\hline
\end{tabular}
\caption{Pay-per-use price (/image) comparison on different  APIs\footnotemark[1]$^,$\footnotemark[2].}
\label{tab:price}
\end{table}

\section{Background and Related Work}
\label{sec:relatedworks}

\subsection{GAN-Based Image Translation}
\label{subsec:relate_GAN}
GAN functions by training two competing models to ultimately learn the unknown true distribution, $P_{\text{data}}(x)$, of the training data $X$.
The generator model $G$, which creates a synthetic image $G(z)$ from a random variable $z$, and the discriminator model $D$, which operates as a binary classifier to distinguish $G(z)$ from true image $x \sim P_{\text{data}}(x)$ are obtained by solving
\begin{equation}
\begin{aligned}
\mathcal{L}_\text{GAN} = 
      \min_G\max_D ~&{\mathbb E}_{x\sim P_{\text{data}}(x)}[{\rm log}D(x)] \\
    &+ {\mathbb E}_{z\sim P_z}[{\rm log}(1-D(G(z)))],
\label{Eq:gan} 
\end{aligned}
\end{equation}
where $P_z$ is a random distribution. 
Upon convergence (\ie, when $G(z)$ approximates $P_{\text{data}}(x)$), $G$ can produce high-quality and photo-realistic samples, rendering it valuable for many image processing tasks. 

By expanding the role of $D$ to differentiate between images from a source domain and those from a target domain {(\eg, image super-resolution or style transfer)}, GAN {and its variants have} gained widespread application in {I2IT}, which is the primary concern of this work. 
{According to \cite{pang2021image}}, I2IT can be divided into supervised and unsupervised based on whether the training samples in the source and the target domain are paired or not. This work considers the widely used supervised framework Pix2Pix \cite{pix2pixhd} and the unsupervised framework CycleGAN \cite{ugatit}. Details of these two frameworks can be found in {Supp.-A}. 

\subsection{Model Extraction Attacks}
\label{subsec:relate_MEA}
As mentioned in Sec.~\ref{sec:Introduction}, by taking the victim model $F_V$ as a labeling oracle, the goal of MEA involves creating an attack model $F_\mathcal{A}$ that emulates the functionalities of the victim model $F_V$. 
{In classification tasks, this goal translates to finding the decision boundary while reducing the query budget under the assumption of attackers' knowledge. 
Strategies include selecting better query samples with reinforcement learning \cite{mea_knockoff} or active learning \cite{mea_activethief}, or even harnessing GAN to generate images that are close to the secret training data of $F_V$ \cite{mea_ES_attack,mea_confidence}. }

It is crucial to note that such strategies cannot be readily transferred to MEAs of I2IT models. In classification tasks, the outputs of the victim model $F_V$ are labels or confidence scores, which inherently carry rich information about the decision boundary of $F_V$. In contrast, I2IT victim models merely return translated images upon query and do not expose their latent embeddings to attackers. Moreover, it remains unknown \cite{mea_gan1,mea_gan_I2IT} what kind of translated images contain more information about the victim $F_V$, and this effect is exacerbated when the attacker's query data distribution deviates from the secret training data of $F_V$ {(\ie, the domain shift problem)}. 
As such, instead of looking for better query samples, we adopt an alternative route that directly mitigates the effect caused by domain shift when extracting the underlying models.

\section{Threat Model}
\label{sec:threat}

\subsection{Adversary's Knowledge}

We consider a victim model $F_V: X \rightarrow Y$ that translates images from a source domain $X$ to a target domain $Y$ has been well-trained on a secret dataset $D_{V} = (D^X_V, D^Y_V)$. The victim $F_V$ has been employed as the backbone to provide API service to remote users. 
The attacker $\mathcal{A}$ is knowledgeable of the general service domains (\eg, translating horse pictures to zebra), but he cannot access the structure, parameters, hyperparameters, and the secret training dataset of $F_{V}$. 
The attacker first constructs his own training dataset $D_\mathcal{A}$ by querying the API service with public samples $x\in D^X_\mathcal{A}$ and collecting labeled pairs $(x, F_V(x))$. Then $\mathcal{A}$ develops a local attack model $F_\mathcal{A}$ that mimics the functionality of $F_V$ by solving 
 \begin{equation}
\label{eq:FA}
\mathop{\text{min}}\limits_{F_\mathcal{A}}~\mathbb{E}_{x \sim p_\mathcal{A}(x)}~d(F_\mathcal{A}(x), F_{V}(x)),
\end{equation}
where $p_\mathcal{A}(x)$ represents the data distribution of $D^X_\mathcal{A}$ and $d$ is a distance metric.

\subsection{Adversary’s Goals}
We consider two typical adversary's goals following the existing literature of MEA 
~\cite{accuracy_fidelity}. 

\noindent \textbf{Functionality Extraction} aims to obtain a local replica that is capable of completing the intended function of the victim model. 
Here, we define the Functional Completion Degree ($R_\text{capability}$) to assess the capability of $F_\mathcal{A}$ in accomplishing the I2IT task. 
$R_\text{capability}$ represents the distance between the region into which the source domain falls after being mapped by the attack model and the target domain. It can be written as  
$$R_\text{capability} = \mathbb{E}_{x\sim p_\text{test}(x)}~d(F_\mathcal{A}(x), Y), $$
where $p_\text{test}(x)$ is the distribution of the test data. It's essential to note that $F_\mathcal{A}$ might even outperform $F_{V}$.

\noindent \textbf{Fidelity Extraction}
is to minimize the discrepancy between the output distribution of the attack model and that of the victim model.
We define the Output Fidelity ($R_\text{fidelity}$) as 
$$R_\text{fidelity}=\mathbb{E}_{x\sim p_\text{test}(x)}~d(F_\mathcal{A}(x), F_{V}(x)). $$ 

It is noted that $R_\text{capability}$ and $R_\text{fidelity}$ can vary independently. In the situation that the attacker lacks access to the target domain $Y$, $R_\text{fidelity}$ is a practical metric to evaluate the effectiveness of MEA.

\begin{figure}[t]
\centering
\includegraphics[width=2.5in]{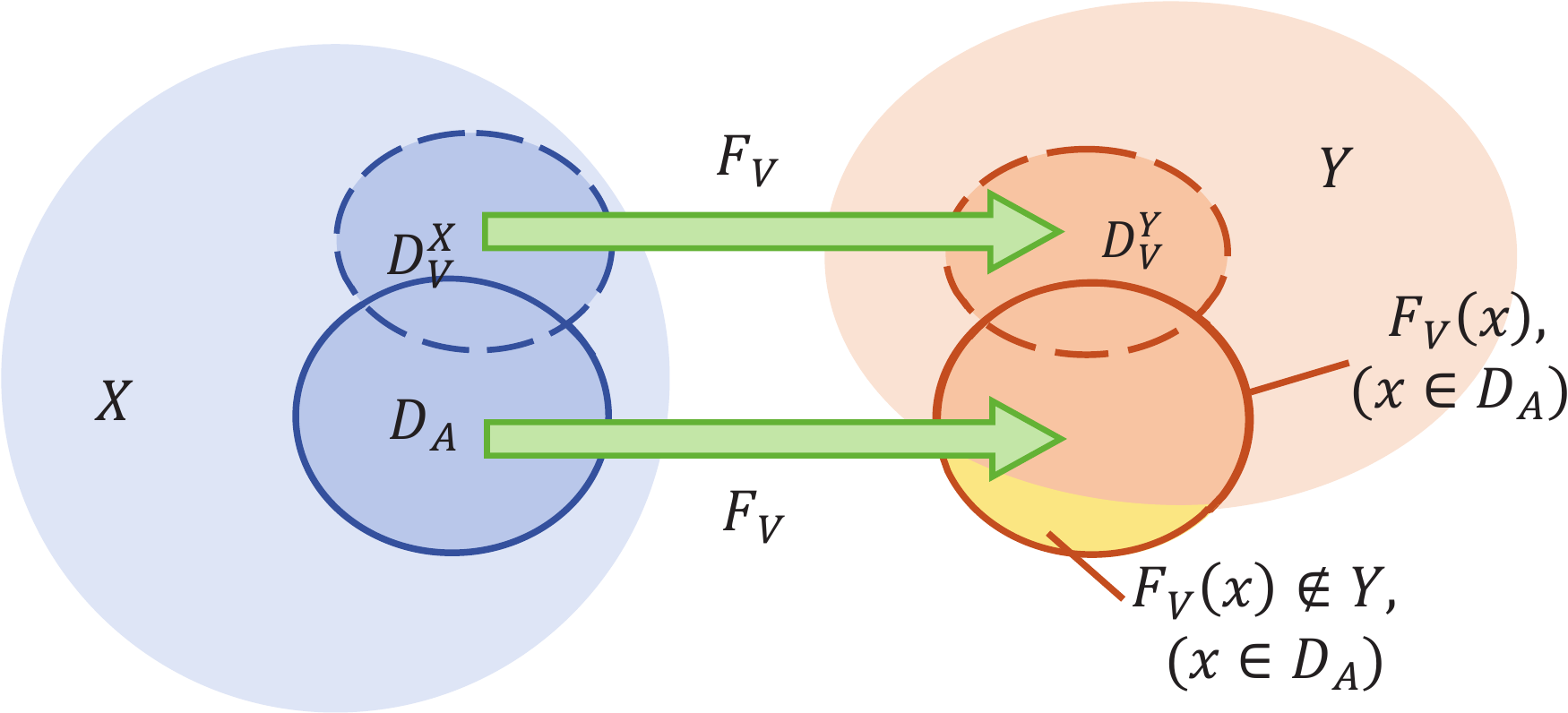}%
\caption{An illustration of the domain shift problem: The problem arises when there is a disparity between the domains of the victim model's training data and the attack data. 
This mismatch causes certain attack data to be incorrectly mapped to the target domain. }
\label{fig:domain shift}
\end{figure}

\begin{figure}[!t]
\centering
    \includegraphics[width=2.5in]{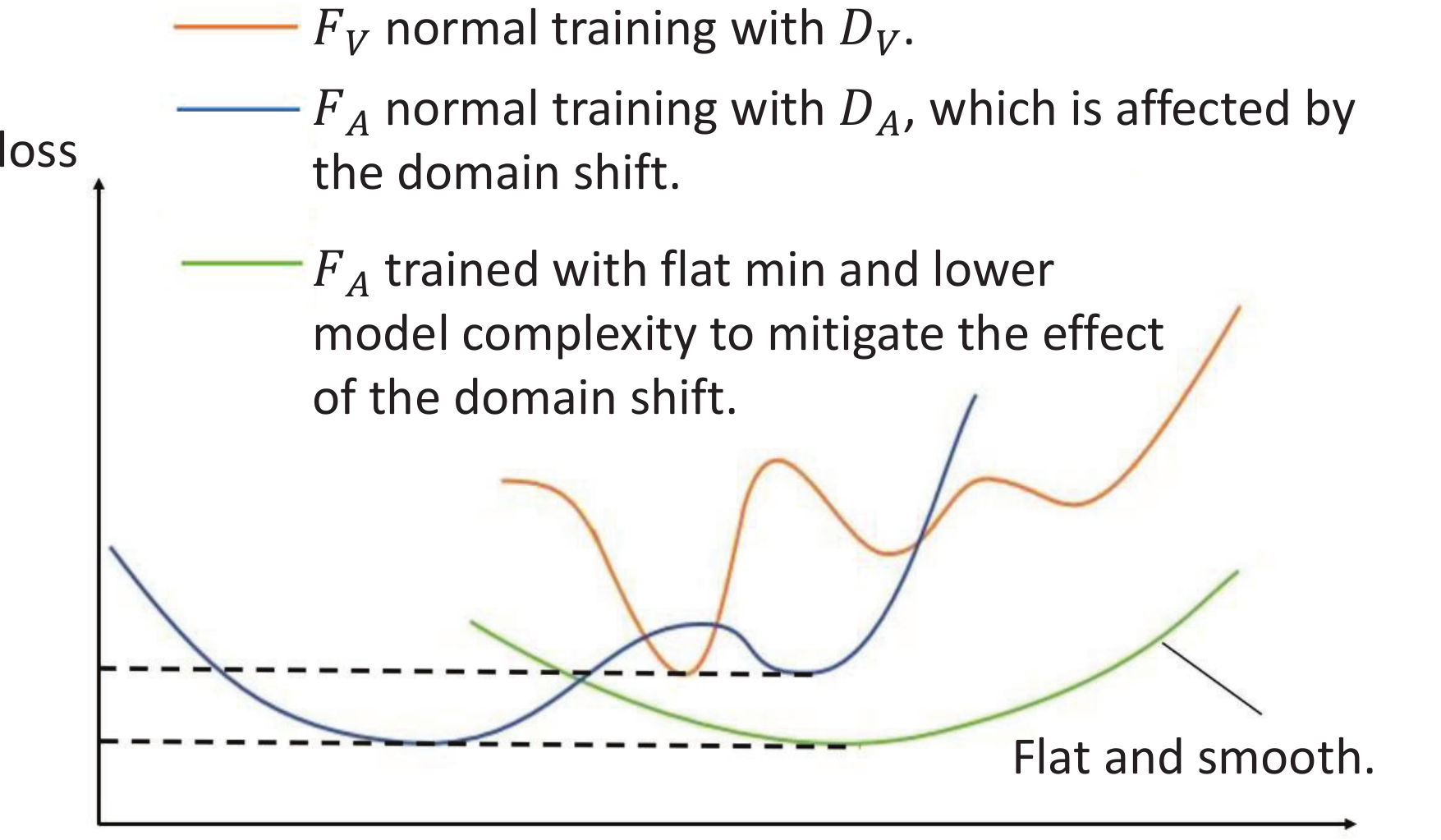}%
    \caption{
    Illustration of the effect of domain shift mitigation. 
    }
    \label{Fig:SAM_Loss}
\end{figure}

\section{Method}
\label{sec:methed}

\subsection{{Domain Shift Problem}}
\label{subsec:Domain Shifting Problem}

The domain shift problem, as discussed in~\cite{domain_shift}, typically pertains to the decline in model performance caused by disparities between the distribution of training data and that of test data. 
In this work,  we identify this phenomenon as a fundamental factor affecting the performance of MEA. 
In MEAs, attackers lack access to the secret data used to train the victim model and they usually only use public datasets as attack data. When a disparity exists between the distribution of this public data and the distribution of the secret data, it causes the attack model's output quality to decrease and can even result in deviations from the target domain. Such data that fails to effectively represent the mapping relationship between the source and target domains can be considered noise data, which hinders the training of subsequent attack models.

Fig.~\ref{fig:domain shift} demonstrates such a scenario. 
The victim model $F_V$ is trained to transfer images from the source domain $X$ to the target domain $Y$. However, the knowledge it has learned is solely based on the data distributions of $D^X_V$ and $D^Y_V$ in its training dataset $D_V$. Consequently, 
the domain shift between $D^X_V$ and $D^X_\mathcal{A}$ can lead to situations where the model's output may not necessarily correspond to the target domain, meaning that $F_V(x)$ might not be a valid representation in $Y$ for images $x \in D^X_\mathcal{A}$.
The situation becomes worse when the attacker utilizes {a publicly available dataset}.
In such cases, the discrepancies between the data distribution in $D^X_V$ and $D^X_\mathcal{A}$ could be significant, exacerbating the challenges of mapping the model's output to the target domain accurately. 
We visualize this discrepancy with real examples from style transfer tasks in {Supp.-B}.

\begin{figure}[t]
\begin{center}
    \includegraphics[width=3.2in]{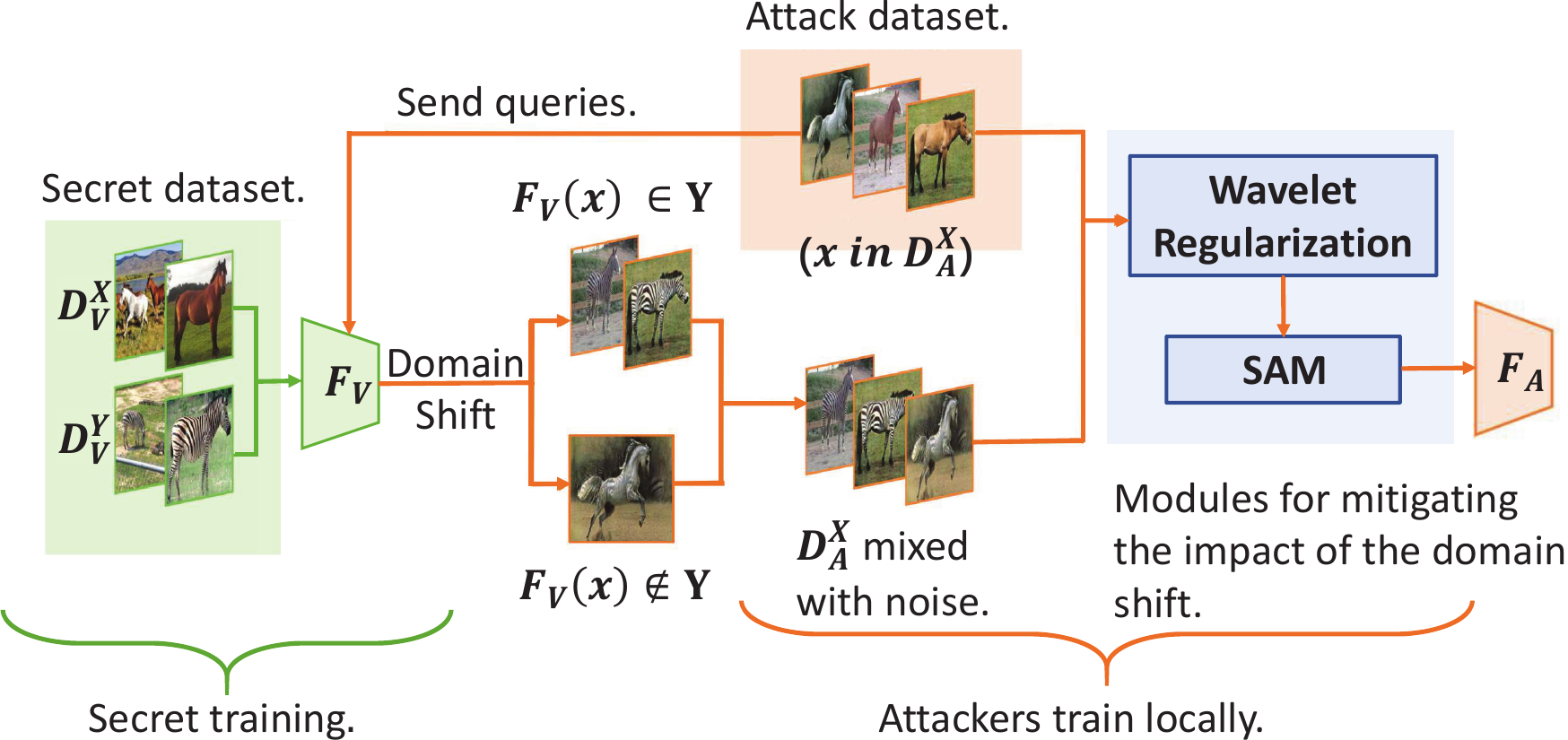}%
    \caption{
    Approach overview.
    }
    \label{Fig:structure}

\end{center}
\end{figure}

\subsection{Attack Overview}
For MEA of classification tasks, traditional wisdom aims at closing the distributional gap between the victim dataset and the attack dataset~{\cite{mea_knockoff,mea_activethief,mea_ES_attack}}.
Due to the unique challenge of I2IT tasks, this work aims to address the impact of the domain shift problem from an orthogonal perspective by resorting to {a flatter and smoother} loss landscape for the attack model.  
Fig.~\ref{Fig:SAM_Loss} illustrates our insight. In the sketch, we briefly illustrate the effect of different factors on the model's test loss. 
{Under vanilla training of the attack model, the obtained $F_\mathcal{A}$ will be fitted to the distribution of the attack's training set $D_\mathcal{A}$, hence deviating from the original optima of $F_{V}$.} In contrast, when the model is enforced to be with lower complexity (\ie, smooth) around a flat minimum area, such local overfitting introduced by domain shift can be mitigated.

To achieve this, we introduce the following two tailored components (as shown in Fig.~\ref{Fig:structure}). 
We first construct a wavelet regularization term from a frequency perspective. 
This concept draws inspiration from recent investigations~\cite{kd_wavelet_gan} which highlight a particular property of GANs' behavior within the frequency domain. Specifically, GANs tend to exhibit low errors within the low-frequency range but usually fail to produce high-quality results in the high-frequency bands. 
We therefore apply the {discrete wavelet transform (DWT)} to decompose images in $D_\mathcal{A}^X$  (i.e., the attackers' input images) and $D_\mathcal{A}^Y$ (i.e., the images generated by the victim model's from the attacker's images) into different frequency bands, and penalize the L1 distance on the high-frequency band. 
The wavelet regularization term can effectively reduce the complexity of the I2IT network. This, in turn, encourages  consistent outputs between $F_V$ and $F_\mathcal{A}$, particularly enhancing finer image details within the high-frequency band. 

We then train the attack model using {sharpness-aware minimization~(SAM)}. 
Due to the noise caused by domain shifting (\ie, $D_V \rightarrow D_\mathcal{A}$), the inherent issue of mode collapse during GAN training can be exacerbated, which leads the model to overly specialize in certain patterns during the generation process, resulting in an overfitting to partial data~\cite{d2020triple}. 
Recent studies have demonstrated that {SAM}~\cite{sam} optimizer has held the promise of seeking out flatter minima by simultaneously minimizing loss value and loss sharpness. 
However, to the best of our knowledge, there have been no prior works reporting the process of training GANs using SAM. 
We therefore introduce a GAN-specific SAM variant towards a wide minimum and further improve the effectiveness of MEA against the I2IT network.

\subsection{Wavelet Regularization}
\label{subsec:Wavelet}
To construct a wavelet regularization term, we first decompose the output image using the DWT into four distinct sub-bands: low frequency ($LL$), low-high frequency ($LH$), high-low frequency ($HL$), and high-high frequency ($HH$). We assume the image intended for DWT as $c$. This can be represented as $\Psi(c) = \{LL(c), LH(c), HL(c), HH(c)\}.$
We denote $\Psi_{p}^H(c) = \{ LH(c), HL(c), HH(c)\}$ as high-frequency terms, where {$p$ represents the times  DWT applied}. 
We then use the L1 distance to construct the wavelet regularization term as
\begin{eqnarray*}
\label{eq:wavelet regularization}
{L}_w^p=\mathbb{E}_{x \sim p_\mathcal{A}(x)}~\left\|\left(\Psi_{p}^H \circ F_\mathcal{A}\right)\left(x\right)-\left(\Psi_{p}^H \circ F_V\right)\left(x\right)\right\|_1.
\end{eqnarray*}
As a result, the overall loss function, denoted as $L$, becomes 
\begin{eqnarray}
L =  L_o +\alpha {{L}_w^p},  
\label{Eq:waveletLoss}
\end{eqnarray}
where $L_o$ is the vanilla loss function of the model backbone (\eg, Pix2Pix or CycleGAN) in I2IT tasks,
and $\alpha$ is the 
coefficient used to balance between $L_o$ and ${L}_w^p$. 

By incorporating ${L}_w^p$ into the training process, our objective is to minimize the disparity in high-frequency information between the victim model's output ($F_V(x)$) and the attack model's output ($F_\mathcal{A}(x)$).  The regularization term can also penalize the model during training, thereby reducing model complexity and alleviating the overfitting to the noise. 

\begin{figure}[t]    
\centering    
\subfloat{    
    \label{fig:loss_landscape:a} 
    \includegraphics[width=1.55in]{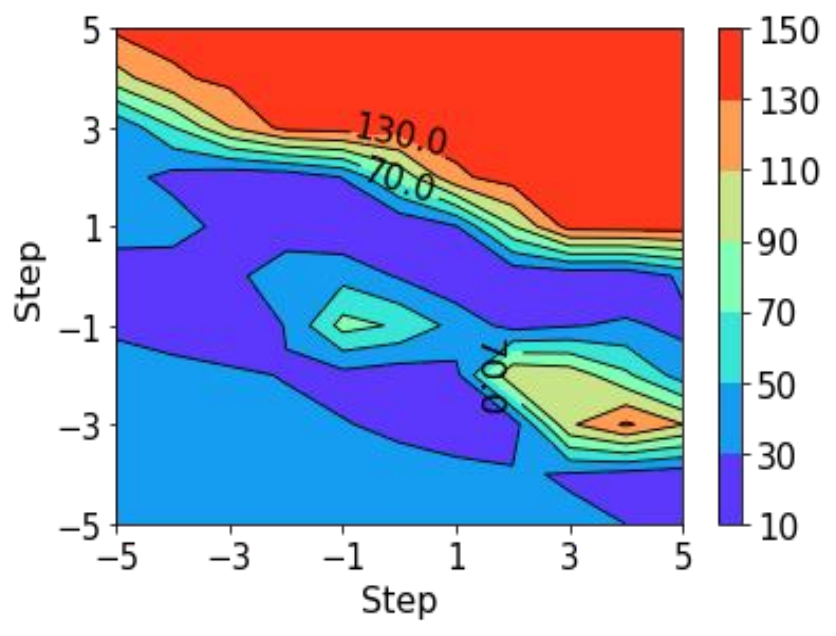}}    
\subfloat{
    \label{fig:loss_landscape:b } 
    \includegraphics[width=1.55in]{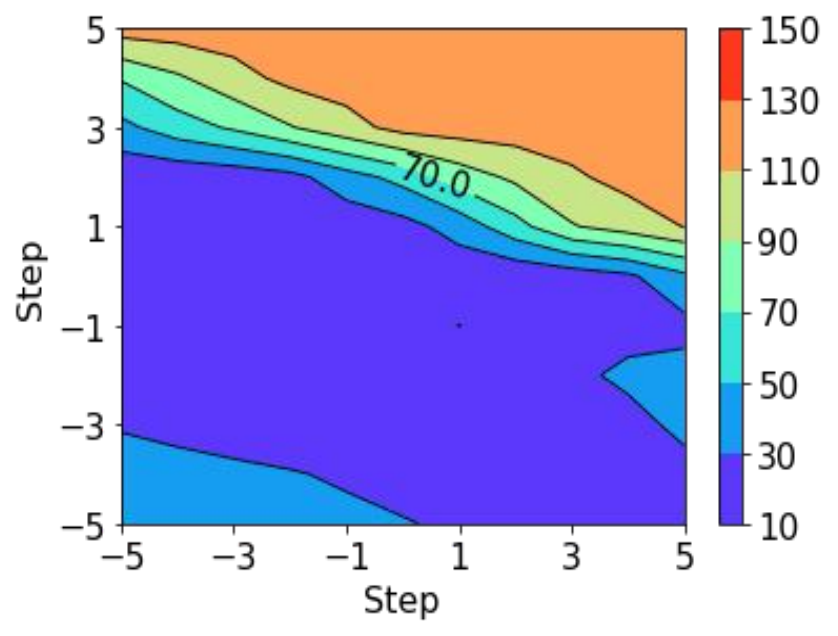}} 
\caption{Comparison of model loss landscape. Left and right are losses trained with Adam and SAM, respectively.
}
\label{fig:loss_landscape} 
\end{figure}

\subsection{SAM for GAN-Based I2IT}
Besides designing a new regularization term (\ie, Eq.~\ref{Eq:waveletLoss}), we also investigate domain shift mitigation from the view of optimizers. 
Assume that the loss function of the model to be optimized is $L$ and the model parameters are $\mathbf{w}$, loss sharpness is defined as
\begin{align*}
\max\limits_{\|\epsilon\|_2\leq\rho}L(\mathbf{w}+\epsilon)-L(\mathbf{w}),
\end{align*}
where $\rho$ ($\rho \ge 0$) is the neighborhood size.
SAM~\cite{sam} optimizes both the loss function and the loss sharpness by solving
\begin{align*}
\min\limits_{\mathbf{w}} \left( \max\limits_{\|\epsilon\|_2\leq\rho}L(\mathbf{w}+\epsilon)-L(\mathbf{w}) \right) +L(\mathbf{w}).
\end{align*}

With the first-order Taylor expansion, the value $\epsilon$ that solves the inner maximization is 
\begin{align*}
\epsilon(\mathbf{w})=\rho\cdot\frac{\nabla_{\mathbf{w}}L(\mathbf{w})}{\|\nabla_{\mathbf{w}}L(\mathbf{w})\|_2}.
\end{align*}
Substitute $\epsilon(\mathbf{w})$ back, the minimization can be approximately solved as 
\begin{align}
\label{Eq:SAMfinal}
\mathbf{w}_{t+1}=\mathbf{w}_{t}-\alpha_{t}\cdot\nabla_{\mathbf{w}}L(\boldsymbol{\mathbf{w}})|_{\mathbf{w}+\epsilon(\mathbf{w})},
\end{align}
where $\alpha_{t}$ is the learning rate at time step $t$. Note that solving Eq.~\ref{Eq:SAMfinal} requires solving $\epsilon(\mathbf{w})$ first, and we use Adam twice in our experiments.



Without loss of generality, consider the backbone model used for I2IT consists of $N$ generators $G = \{G_1, ..., G_N\}$ and $M$ discriminators $D = \{D_1, ..., D_M\}$, with their respective parameters being $\mathbf{w}^{G}=\left\{\mathbf{w}^{G_1}, \ldots,\mathbf{w}^{G_N}\right\}$ and $\mathbf{w}^{D} = \left\{\mathbf{w}^{D_1}, \ldots,\mathbf{w}^{D_N}\right\}$. Referring to Eq.~\ref{Eq:gan}, we alternatively optimize $G$ and $D$ with SAM optimizer by  
\begin{equation*}
\left\{
\begin{aligned}
&\epsilon(\mathbf{w}^{G_i})=\rho_{G_i}\cdot\frac{\nabla_{\mathbf{w}^{G_i}}L(\mathbf{w})}{\|\nabla_{\mathbf{w}^{G_i}}L(\mathbf{w})\|_2}, \\
&g(\mathbf{w}^{G_i})=\nabla_{\mathbf{w}^{G_i}}L(\mathbf{w}^{G_i})|_{\mathbf{w}^{G_i}+\epsilon(\mathbf{w}^{G_i})},\\
&\mathbf{w}_{t+1}^{G_i}=\mathbf{w}_t^{G_i}-\alpha_{t}\cdot g(\mathbf{w}^{G_i}),
\end{aligned}
\right.
\end{equation*}
for each $G_i\in G~(i \in [1, N])$ and
\begin{equation*}
\left\{
\begin{aligned}
&\epsilon(\mathbf{w}^{D_j})=\rho_{D_j}\cdot\frac{\nabla_{\mathbf{w}^{D_j}}L(\mathbf{w})}{\|\nabla_{\mathbf{w}^{D_j}}L(\mathbf{w})\|_2},\\
&g(\mathbf{w}^{D_j})=\nabla_{\mathbf{w}^{D_j}}L(\mathbf{w}^{D_j})|_{\mathbf{w}^{D_j}+\epsilon(\mathbf{w}^{D_j})},\\
&\mathbf{w}_{t+1}^{D_j}=\mathbf{w}_t^{D_j}+\alpha_{t}\cdot g(\mathbf{w}^{D_j}), 
\end{aligned}
\right.
\end{equation*}
for each $D_j\in D~(j \in [1, M])$. Here, $\rho_{G_i}$ and $\rho_{D_j}$ are the loss sharpness hyper-parameters of $G_i$ and $D_j$, respectively.

We summarize the process in algorithm in the Supp.-C. It is imperative to highlight that the scenario under consideration resembles that of the CycleGAN model, {where two generators and two discriminators are used and the generators iterate concurrently}. However, if certain generator parameters require individual iteration, they should have respective optimization directions computed separately. As gradients need to be computed twice, the SAM requires forward propagation to be performed twice in each iteration.

We conduct experiments in the horse2zebra task, and visualize the loss landscape of the Pix2Pix backbone by training models 
with different optimizers in Fig.~\ref{fig:loss_landscape}.
After applying SAM, the loss landscape of $F_A$ becomes flatter.

\section{Experimental Analyses}
\label{sec:Experiments}

\begin{figure}[t!]
\begin{center}
	\includegraphics[width=0.45\textwidth]{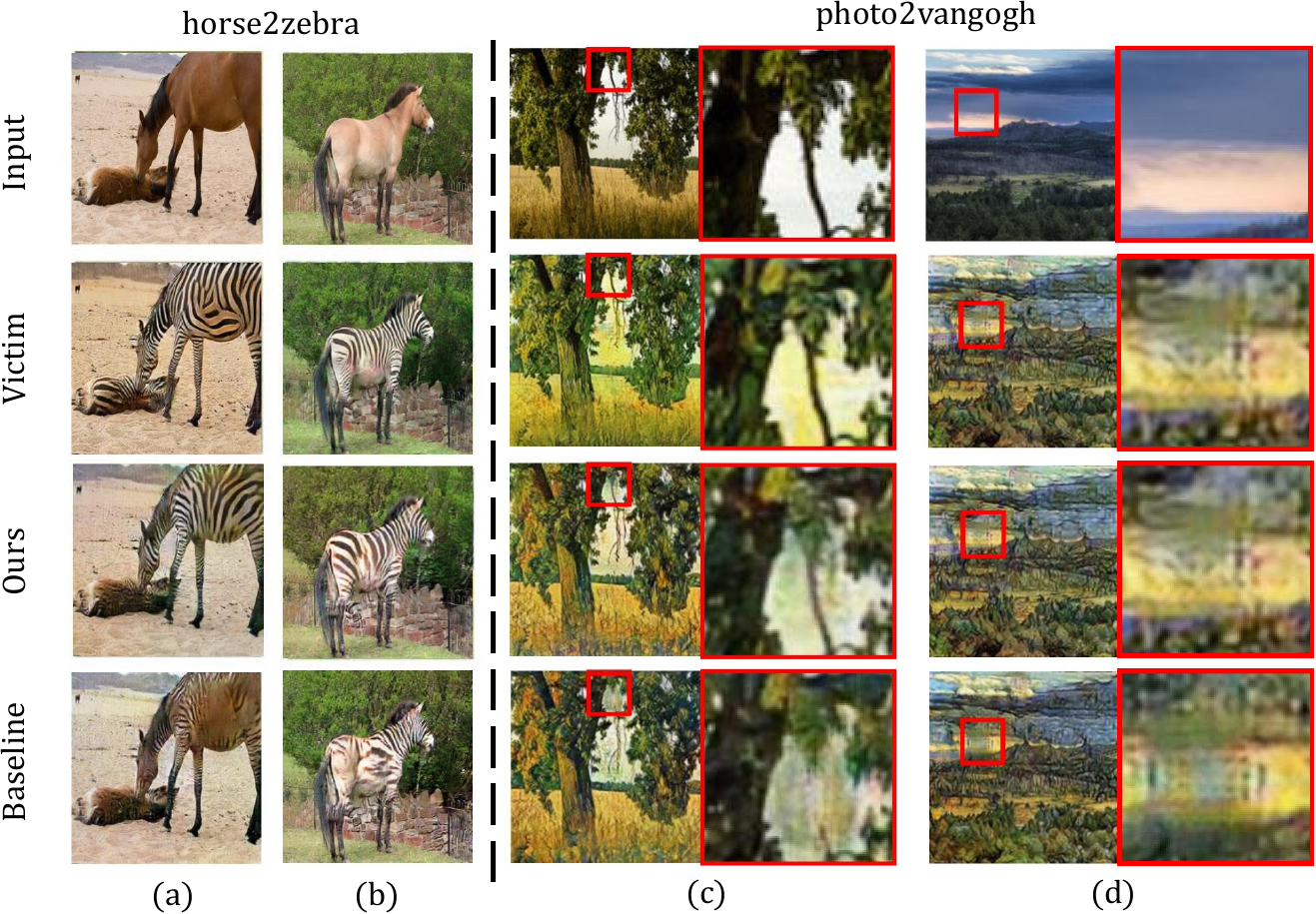}
	\caption{Qualitative results of our attack against the style transfer task. (a-b). horse2zebra, (c-d). photo2vangogh.}
	\label{fig:result_duibi}
\end{center}
\end{figure}

\subsection{Experimental Setup}
\par\noindent\textbf{Dataset and Models}.
We assess the performance of our attack on typical I2IT tasks, \ie, style transfer tasks including horse2zebra (converting horse images to zebras) and photo2vangogh (converting photos to Van Gogh style), as well as the super-resolution task (enhancing anime images resolution).

To build the victim model, we use CycleGAN~\cite{cyclegan} to train the style transfer tasks, and we directly employ the pre-trained real-ESRGAN model~\cite{Real-esrgan} for the super-resolution task. 
The datasets employed to train the victim style transfer models are horse2zebra and photo2vangogh in~\cite{CycleGAN_datasets}. The datasets used for training the victim super-resolution model {are DIV2K~\cite{div2k}, Flickr2K~\cite{Flicker2K_datasets}, and OutdoorSceneTraining~\cite{dataoutdoor-scene-training_datasets}}. All of these are identical to the original settings~\cite{cyclegan,Real-esrgan}. All victim models have demonstrated compelling performance in their respective tasks.  

To train the model, we use Pix2Pix~\cite{pix2pix} and CycleGAN~\cite{cyclegan} as attack modeling backbone frameworks for style transfer and super-resolution tasks, respectively.

Since attacker $\mathcal{A}$ is knowledgeable of the general service domains, we construct $\mathcal{A}$'s training dataset $D_\mathcal{A}^X$ as follows. 
For the horse2zebra task, we employ horse images from the Animal10 dataset~\cite{horse_datasets}. For the photo2vangogh task, we utilize a subset of two thousand landscape images from the Landscape dataset~\cite{Landscape_datasets}. Regarding the super-resolution task, we use the Anime dataset~\cite{Hayao_and_Shinkai_datasets}, which comprises images sourced from anime films created by Makoto Shinkai, Hayao Miyazaki and Kon Satoshi. 

We employ the test set of the victim model to evaluate the performance on style transfer tasks.  We select 200 images from Anime dataset~\cite{Hayao_and_Shinkai_datasets} (ensuring they are disjoint with the training dataset used for the attack) as the test set for the super-resolution task. Further details about the setup can be found in Supp.-D.

\par\noindent\textbf{Evaluation Metrics}.
We assess the image quality of our attack using widely adopted metrics: Fréchet Inception Distance (FID) \cite{fid} and Kernel Inception Distance (KID) \cite{kid}. Both FID and KID measure the divergence between image distributions, with lower scores implying higher similarity. Additionally, we employ PSNR (Peak Signal-to-Noise Ratio) and LIPIPS \cite{lpips} at both pixel and perceptual levels. PSNR calculates pixel-wise differences between images, with higher values denoting greater similarity. Whereas, LPIPS incorporates perception-based features, with lower values indicating increased similarity.

\par\noindent\textbf{Baseline}. 
We consider the Artist-Copy~\cite{mea_gan_I2IT} as the baseline since it is currently the only known MEA in GAN-based I2IT.

\begin{figure}[t]
\begin{center}
	\includegraphics[width=0.45\textwidth]{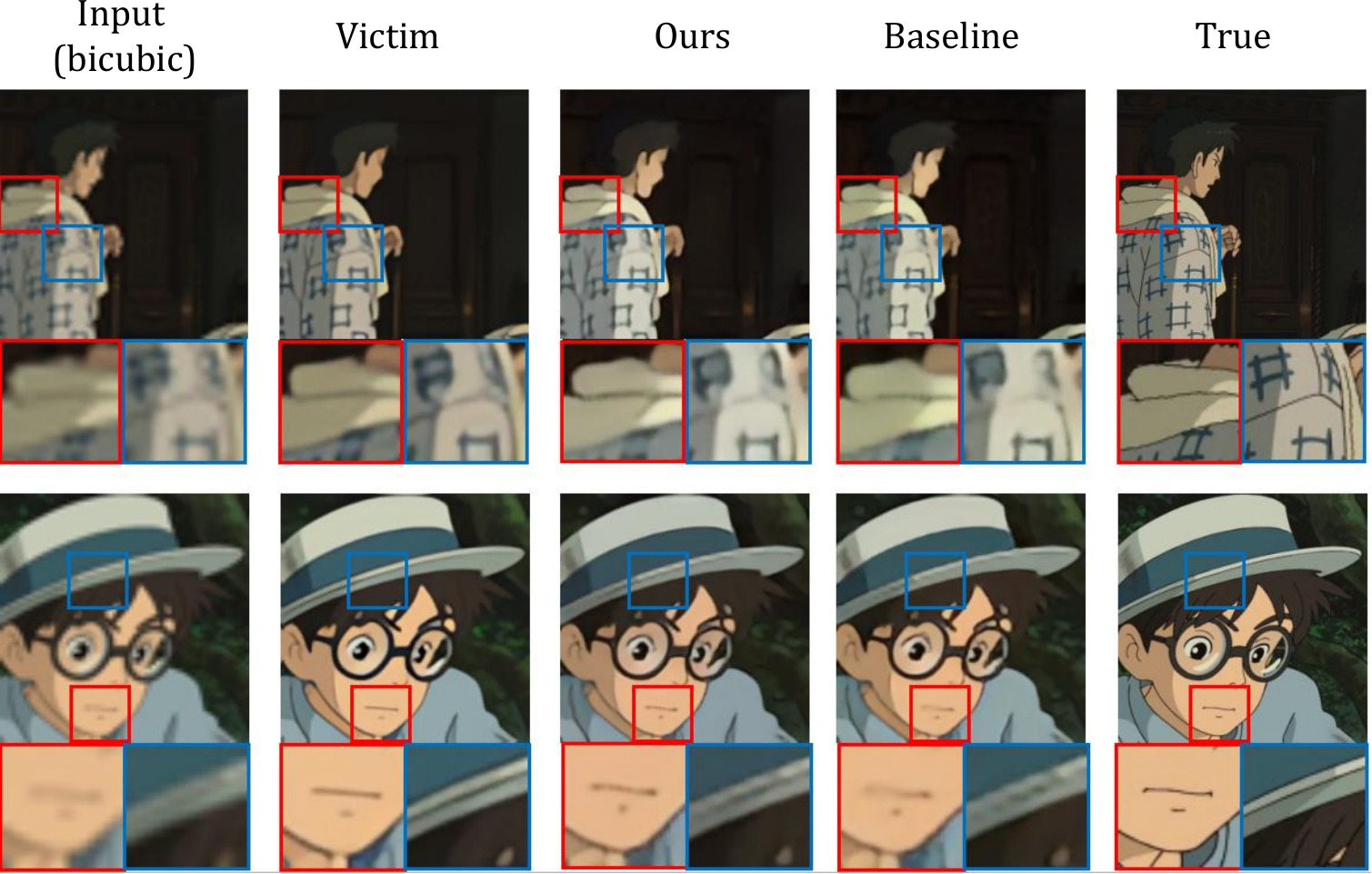}
	\caption{Qualitative results of our attack against the super-resolution task (4x upscaling).}
	\label{fig:chaofen_duibi}
\end{center}
\end{figure}

\begin{table}[t]
\centering
\fontsize{9pt}{10pt}\selectfont
\begin{tabular}{| c | c | c |c|c | c |}
\hline
    \multirow{2}{*}{Task}& \multirow{2}{*}{Method} 
    &\multicolumn{2}{|c|}{$R_\text{capability}$} &\multicolumn{2}{|c|}{$R_\text{fidelity}$} \\
\cline{3-6}
     &  
     & FID$\downarrow$ & KID$\downarrow$ & FID$\downarrow$ & KID$\downarrow$ \\
\hline
        \multirow{3}{*}{\centering h2z\textsuperscript{$\dagger$}} &Victim   
        & 63.08 & 1.21±0.09 & NA & NA  \\
\cline{2-6}
        & A-C 
        & 115.09 & 5.59±0.39 & 75.65&1.93±0.25 \\
\cline{2-6}
        &Ours 
        & \textbf{82.63} & \textbf{2.55±0.24} & \textbf{57.87}& \textbf{0.32±0.09}  \\
\hline
        \multirow{3}{*}{\centering p2v\textsuperscript{$\dagger$}} &Victim   
        & 109.78 & 2.73±0.43 &NA & NA \\
\cline{2-6}
        ~& A-C
        & 112.31 & 3.01±0.41 & 59.83 & 1.70±0.27\\
\cline{2-6}
        ~&Ours 
        & \textbf{110.00} & \textbf{2.66±0.37}& \textbf{59.09} &\textbf{1.45±0.25}  \\
\hline
\end{tabular}
\begin{flushleft}
\scriptsize
\textsuperscript{$\ddagger$} Since the consistency between the victim model's output and itself is not required, the $R_\text{fidelity} $ values are represented as NA. KID is reported as KID×100±std.×100, the same hereinafter. 
\textsuperscript{$\dagger$} We use h2z and p2v to represent horse2zebra and photo2vangogh. 
\end{flushleft}
\caption{\label{tab1}Quantitative results of our attack against the style transfer tasks\textsuperscript{$\ddagger$} (A-C represents Artist-Copy).} 
\end{table}

\subsection{Attack Performance}

\subsubsection{Style Transfer.}
Compared to the baseline, our method demonstrates significant performance improvement in all comparative experiments. 

As shown in Table~\ref{tab1}, in the horse2zebra task, the {FID/KID} scores of our attack  reach {82.63/2.55 for $R_\text{capability}$ and 57.87/0.32 for $R_\text{fidelity}$}, significantly surpassing the Artist-Copy method which achieves 115.09/5.59 and 75.65/1.93 FID/KID scores for $R_\text{capability}$ and $R_\text{fidelity}$, respectively. 
In the photo2vangogh task, our attack achieves {110.00/2.66 and 59.09/1.45} FID/KID scores for $R_\text{capability}$ and $R_\text{fidelity}$. In contrast, the Artist-Copy method only attains FID/KID values of 112.31/3.01 for $R_\text{capability}$ and 59.83/1.70 for $R_\text{fidelity}$. 
 
It is worth mentioning that in the horse2zebra task, the attack model shows the highest improvement, with a decrease of 32.46 in $R_\text{capability}$ FID. Moreover, in the photo2vangogh task, our model achieves performance on $R_\text{capability}$ that is almost as good as the victim model.

Fig.~\ref{fig:result_duibi} demonstrates the effect of our attack in style transfer tasks. 
As shown in Fig.~\ref{fig:result_duibi} (a-b), our attack successfully extracts the functionality of the victim model, enabling the transformation of horses into zebras while 
the Artist-Copy method struggles to properly add zebra patterns to the horses in the horse2zebra task. 
Fig.~\ref{fig:result_duibi} (c-d) illustrates the attack performance against the photo2vangogh task with zoomed-in views of the outputs in order to demonstrate our attack's capacity in generating fine details of the images. 
We can observe that our attack has achieved closer outputs to the victim model and it  handles the details better due to its ability in generating higher-quality high-frequency information. 
However, the Artist-Copy method shows significant abnormal noise and artifacts in the output due to overfitting on noisy data induced by domain shift problems. Additional results can be found in Supp.-E.

Notably, our attack necessitates a significantly smaller number of queries  compared to the baseline. 
For example, the performance of our attack using half the number of queries (2-3k queries) surpasses that of the baseline (4-5k queries). Results can be found in Supp.-E).
These findings highlight the impact of the domain shift issue, a factor that significantly limits the effectiveness of traditional MEAs which rely on increasing query counts to enhance attack capabilities. In contrast, our MEA, incorporating methods that address the domain shift problem, achieves favorable performance with a substantially reduced number of queries.  

\subsubsection{Super-resolution.}
Our method also achieves better attack performance in the super-resolution task (4x upscaling). In Table~\ref{tab:super-resolution}, we present the experimental results. 
The metrics in the table are computed as the averages of the test set images. 

Compared to the Artist-Copy, our method shows a 2.06 increase in PSNR for $R_\text{capability}$ and a decrease of 0.032 in LIPIS. Moreover, our method exhibits a 2.16 increase in PSNR for $R_\text{fidelity}$ and a decrease of 0.015 in LIPIS.

In Fig.~\ref{fig:chaofen_duibi}, we compare the performance of bicubic interpolation (i.e.,  a common method for image upscaling), victim, Artist-Copy, and our method in the super-resolution task. 
We can observe that our method's output shows a significant improvement in image clarity compared to the bicubic method, demonstrating the success of the attack. Furthermore, through an analysis of the performance on the detailed parts of Fig.~\ref{fig:chaofen_duibi}, it can be noted that our method's output presents texture details with greater clarity compared to the Artist-Copy. For example, in the second row of Fig.~\ref{fig:chaofen_duibi}, it becomes evident that our method's output exhibits significantly sharper edges on the hat and more distinct lines around the mouth. This observation further implies that our approach consistently aligns with the performance of the victim model, particularly in capturing high-frequency elements within the frequency domain. 

\begin{table}[t]
\centering
\fontsize{9pt}{10pt}\selectfont
\begin{tabular}{| c | c | c |c | c | }
\hline
    \multirow{2}{*}{Method}   &\multicolumn{2}{|c|}{$R_\text{capability}$}  &\multicolumn{2}{|c|}{$R_\text{fidelity}$}\\
\cline{2-5}
     &  PSNR$\uparrow$ &  LPIPS$\downarrow$ & PSNR$\uparrow$ &  LPIPS$\downarrow$ \\
\hline
        Victim   & 34.56& 0.067&NA&NA\\
\hline
        Artist-Copy   & 27.67&0.177 &27.89&0.156\\
\hline        
        Ours& \textbf{29.73} &\textbf{0.145}& \textbf{29.73} &\textbf{0.141}\\
\hline
\end{tabular}

\caption{Quantitative results of our attack against the super-resolution task (4x upscaling).} 
\label{tab:super-resolution}
\end{table}

\subsection{Ablation Study}
We now evaluate the effectiveness of the wavlet regularization and SAM, respectively, in model extraction attacks against the horse2zebra task. Table~\ref{tab:Ablation Experiment} shows that 
both the FID and KID scores of $R_\text{capability}$ and $R_\text{fidelity}$ undergo a significant increase when either the SAM or the wavelet regularization term is removed, 
which demonstrates the necessity and effectiveness of each component in our method.

The analysis reveals that the use of the wavelet regularization term alone on top of the baseline gives a better performance in terms of both $R_\text{fidelity}$ and $R_\text{capbility}$ because it promotes the consistency of the outputs of the victim model and the attack model in the frequency domain. 
On top of it, SAM can further improve the attack performance due to its ability in 
finding a much flatter optimum. 

\begin{table}[t]
\begin{center}
\fontsize{9pt}{10pt}\selectfont
\begin{tabular}{| c | c | c |c | c |c |}
\hline
    \multicolumn{2}{|c|}{Component}&\multicolumn{2}{|c|}{$R_\text{capability}$} &\multicolumn{2}{|c|}{$R_\text{fidelity}$}  \\
\hline
    ${L}_w^p$ & SAM & FID$\downarrow$ & KID$\downarrow$ & FID$\downarrow$ & KID$\downarrow$\\
\hline
    $\times$ & $\times$&115.09 & 5.59±0.39 & 75.65&1.93±0.25\\
\hline
    $\surd$ & $\times$&100.69 & 4.25±0.30 &68.01  &1.06±0.17\\
\hline
    $\times$ & $\surd$&104.04 &  4.37±0.30&  70.72 &1.25±0.22\\
\hline
    $\surd$ & $\surd$&\textbf{82.63} & \textbf{2.55±0.24} & \textbf{57.87}& \textbf{0.32±0.09}\\
\hline
\end{tabular}
\end{center}
\caption{\label{tab:Ablation Experiment}Ablation Experiment Results.}
\end{table}

\section{MEA Agaisnt I2IT in Real Life}
\label{sec:real}

\begin{figure}[t!]
\begin{center}
  \includegraphics[width=3.2in]{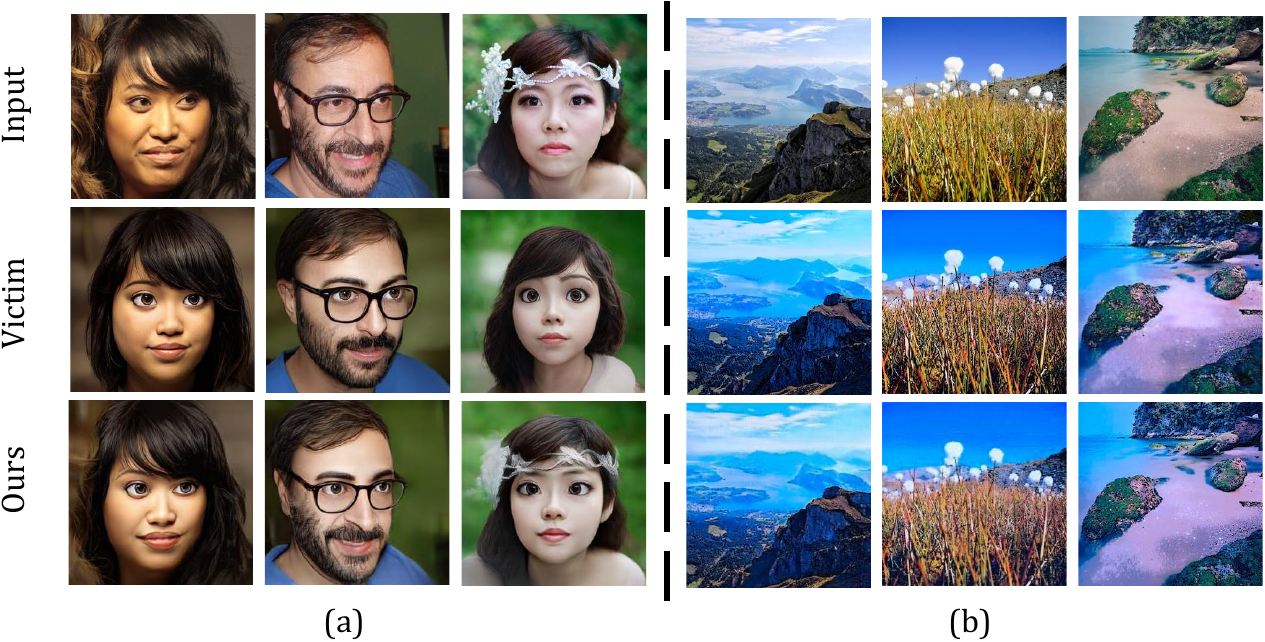}
  \caption{The performance of our attack against commercial
I2IT services. (a). face2cartoon (3d), (b). landscape2cartoon.} 
  \label{fig:real-shili}
\end {center}
 \end{figure}

\begin{table}[t!]
\begin{center}
\fontsize{9pt}{10pt}\selectfont
\begin{tabular}{|c|c|c|c|}
\hline
    Task   
    & Method & FID$\downarrow$ & KID$\downarrow$ \\
\hline
    \multirow{2}{*}{\centering face2cartoon (2d)}    & Artist-Copy &  63.20 & 1.79±0.27 \\
\cline{2-4}
       ~& {Ours} & \textbf{59.61} &\textbf{1.36±0.2} \\
\hline
    \multirow{2}{*}{\centering face2cartoon (3d)}   & Artist-Copy & 60.89 & 1.43±0.18\\
\cline{2-4}
        & {Ours} & \textbf{51.67} & \textbf{0.61±0.13}\\
\hline
    \multirow{2}{*}{\centering landscape2cartoon} & Artist-Copy & 68.55 & 1.47±0.21\\
\cline{2-4}
        & {Ours} & \textbf{64.65} & \textbf{1.21±0.20}\\
\hline
\end{tabular}
\end{center}
\begin{flushleft}
\end{flushleft}
\caption{\label{tab:real}The performance of our attack against commercial I2IT
services.}
\end{table}

In this section, we conduct MEA against real-world commercial I2IT services to verify our attack's ability. 
We select two widely used I2IT service platforms as our target. 
The outcomes of our attack substantiate its impressive efficacy in successfully extracting victim models.

\subsection{I2IT Services}
The mainstream I2IT services can be divided into two categories: style transfer, which facilitate the conversion of input image styles, and image enhancement services, which focus on restoring and improving the quality of degraded images.

Typically, users can access the I2IT services through two primary ways. The first involves utilizing an API in the cloud, following a pay-as-you-go model. The other is to access the model features locally through buyout purchases, allowing for an unlimited number of interactions. 
Nonetheless, the local model is typically encapsulated or encrypted within a black box, which constrains users from accessing its internal components, such as the model's architecture and parameters. 
In both scenarios, MEAs can be performed by sending queries to the black-box target model. 

\subsection{Attack Process}
\paragraph{Attack Settings}
We select two popular third-party I2IT service providers in the market:  Imglarger\footnotemark[2] and Baidu AI Cloud\footnotemark[1]. We conduct MEAs against Imglarger's  human face cartoonization functions, including face2cartoon (2d) and face2cartoon (3d), as well as Baidu AI Cloud's picture style cartoonization function (i.e., landscape2cartoon). 

We randomly select 2,000 images from the face dataset FFHQ (512$\times$512)~\cite{dataset_ffhq} and Landscape dataset~\cite{Landscape_datasets} as the attack dataset, respectively. The face dataset FFHQ is composed of 52,000 high-quality PNG images with 512$\times$512 resolution crawled from Flickr. These images vary considerably in age, race, and image background, and are automatically aligned and cropped using dlib. We use the images in this dataset as the attack dataset for both face2cartoon (2d) and face2cartoon (3d) tasks and choose CycleGAN as the attack model backbone. The Landscape dataset is the same we used in the photo2vangog task, 
and Pix2Pix is used as the attack model backbone for landscape2cartoon task. To evaluate the extraction performance, we take 200 and 150 images from the remaining parts of the FFHQ and landscape datasets, respectively, as the test set.

\subsection{Results Comparison}
We now analyze the results in both quantitative and qualitative ways. 
Since the $D_V$ cannot be obtained from the commercial services, we cannot measure the distance from the attack model output to the target domain (\ie, the $R_\text{capability}$). 
As a result, we only evaluate the attack model using the $R_\text{fidelity}$ metric in this section. 
Fig.~\ref{fig:real-shili} and Table~\ref{tab:real} illustrate the performance of our attack. The results show that our approach successfully replicates the function of the victim model, surpassing the performance of the baseline method. Additional results can be found in Supp.-E.

\section{Conclusion}
In this paper, we have presented a novel MEA attack to extract models in I2IT tasks. 
We identify that the impact of the domain shift problem is a fundamental factor affecting the performance of MEA. 
This is particularly notable for GAN-based I2IT tasks where the optimal selection of queries is not apparent. 
Our approach addresses the issue from a new angle by resorting to a flatter and smoother loss landscape for the attack model.  
By incorporating wavelet regularization and sharpness-aware minimization, our attack exhibits significant performance improvement in all comparative experiments. We also conduct our attack against real-world commercial I2IT services. The outcomes of our attack substantiate its impressive efficacy in successfully
extracting victim models.

Future work will include further investigation on dedicated defense mechanisms. We also see new research opportunities in 
extending the attack approach against diverse GAN-based models.

\section*{Ethics Statement}
{Given the nature of Sec.~\ref{sec:real} which involves discussing real-world attacks, we do not publicly release our code of this section in the usual manner. Instead, we intend to make it accessible only to legitimate researchers as per request. This approach aims to facilitate the reproducibility of our findings while also maintaining a responsible handling of potentially sensitive information.}

\section*{Acknowledgments}
{
This work was partially supported by the National Natural Science Foundation of China (No. 12271464) and the Hunan Provincial Natural Science Foundation of China (No. 2023JJ10038). 

Correspondence of this work should be directed to L. Zhang (leo.zhang@griffith.edu.au) and H. Yuan (yhz@xtu.edu.cn).
}

\bibliography{aaai24}

\clearpage

\section*{Supplementary}

\subsection{A.~~The Details of Pix2Pix and CycleGAN}

\textbf{Pix2Pix} and its variants~\cite{pix2pixhd} 
serve as a classic backbone for image-to-image translation tasks based on Conditional GAN (cGAN). 
Just like traditional GAN, cGAN consists of a generator $G$ and a discriminator $D$, but both $G$ and $D$ also receive an additional input image $x$ as the conditions to facilitate a more profound understanding of the input-output relationship.  Unlike GANs that primarily distinguish between real and synthetic images, the discriminator $D(x, y)$ in cGANs evaluates two aspects simultaneously: whether the image $y$ appears authentic and whether it corresponds to the condition $x$. The cGAN loss function is 
\begin{align*}
\min_G\max_D~& \mathcal{L}_\text{cGAN}(G,D),
\end{align*}
\begin{align*}
\mathcal{L}_\text{cGAN}(G,D) = ~& \mathbb{E}_{x,y}[\log D(x,y)]+ \notag\\
&\mathbb{E}_{x,z}[\log(1-D(x,G(x,z))],
\end{align*}
where $x$ and $y$ are paired samples and $z$ is a random variable. The overall Pix2Pix loss function integrates the cGAN loss and a perceptual loss term $\mathcal L_1(G)$:
\begin{align*}
\mathcal{L}_\text{Pix}(G,D) = \operatorname*{min}_G\operatorname*{max}_D\mathcal ~ L_\text{cGAN}(G,D)+\lambda\mathcal L_1(G),  
\end{align*}
where $\mathcal{L}_1(G)=\mathbb{E}_{x,y,z}[\|y-G(x,z)\|_1]$ promotes the generation of images with reduced blurriness.

\noindent \textbf{CycleGAN}~\cite{ugatit} is an unsupervised framework used for image-to-image translation tasks.
In contrast to Pix2Pix, which necessitates paired data for training, CycleGAN can learn mappings from the source domain $X$ to the target domain $Y$ using cycle-consistency loss from unpaired data.
It employs two sets of generators and discriminators: generator $G_1: X \rightarrow Y$ and its discriminator $D_Y$;  generator $G_2: Y \rightarrow X$ and its discriminator $D_X$.
The overall loss function of CycleGAN is formulated as: 
\begin{align*}
\mathcal{L}_\text{Cycle}\left(G_1, G_2, D_X, D_Y\right) &=\operatorname*{min}_{G_1, G_2}\operatorname*{max}_{D_X, D_Y}\mathcal ~\lambda \mathcal{L}_{\text{cyc}}(G_1, G_2)\notag\\
&+\mathcal{L}_{\text{GAN}}\left(G_1, D_Y, X, Y\right)\notag\\
&+\mathcal{L}_{\text{GAN}}\left(G_2, D_X, Y, X\right),
\end{align*}
where $\mathcal{L}_{\text{cyc}}$ is the cycle consistency loss given by
\begin{align*}
\mathcal{L}_{\text{cyc}}(G_1, G_2)=\mathbb{E}_{x \sim p_\text{data}(x)}\left[\|G_2(G_1(x))-x\|_1\right] \notag\\
+\mathbb{E}_{y \sim p_\text{data}(y)}\left[\|G_1(G_2(y))-y\|_1\right],
\end{align*}
and $\mathcal{L}_{\text{GAN}}\left(G_1, D_Y, X, Y\right)$  is the GAN adversarial loss given by ($\mathcal{L}_{\text{GAN}}\left(G_2, D_X, Y, X \right)$ is similarly defined):
\begin{align*}
\mathcal{L}_{\text{GAN}}\left(G_1, D_Y, X, Y\right)=\mathbb{E}_{y \sim p_\text{data}(y)}\left[\log D_Y(y)\right] \notag\\
+\mathbb{E}_{x \sim p_\text{data}(x)}\left[\log \left(1-D_Y(G_1(x))\right]\right.,
\end{align*}
where $p_\text{data}(x)$ and $p_\text{data}(y)$ are the underlying distributions for domain $X$ and domain $Y$, respectively. 

The cyclic consistency implies that  when an image from one domain is translated into another domain and subsequently translated back (\eg, $x\rightarrow y \rightarrow x'$), the resultant image should align perfectly with the original image. The adversarial losses assess whether the images generated by $G_1$ and $G_2$ appear realistic.

\begin{figure}[ht]    
    \centering
    \subfloat[Victim's training dataset.]{    
        \label{fig:v_train} 
        \includegraphics[width=3.2in]{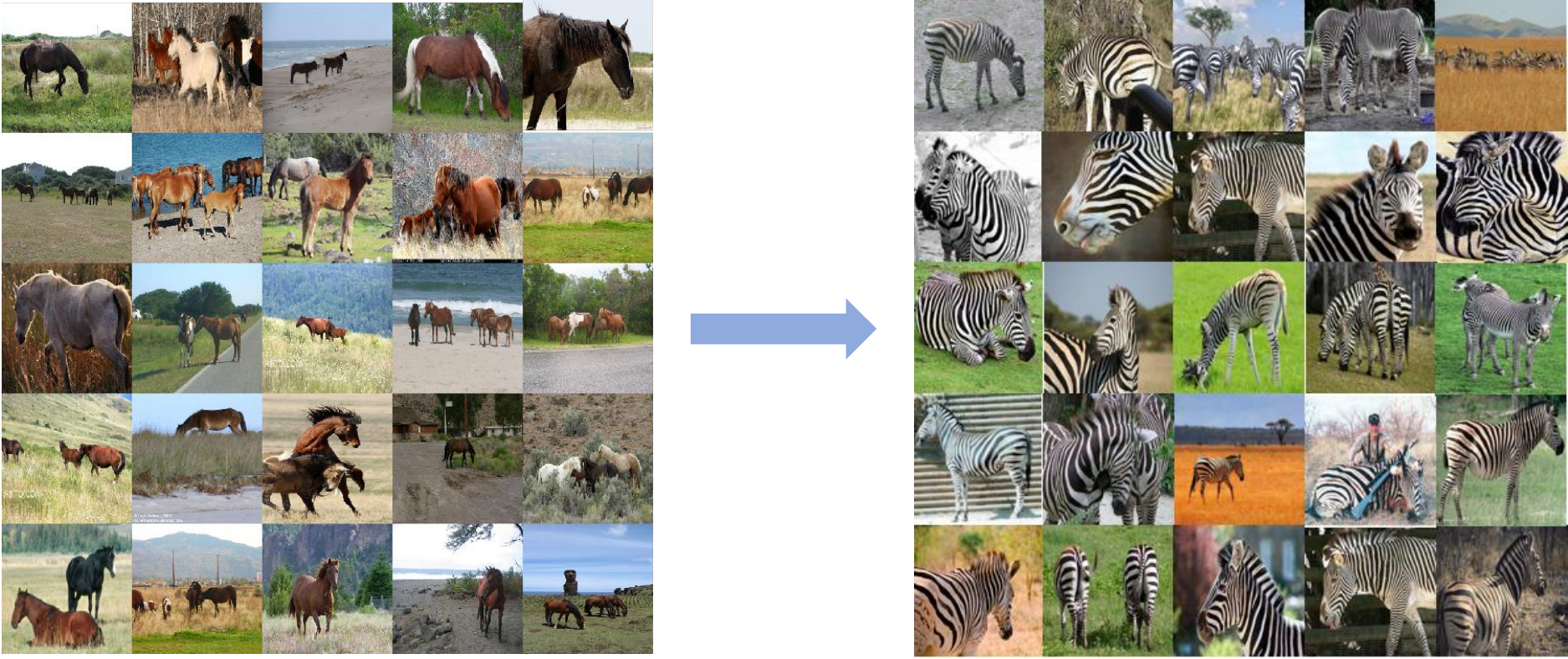}} \\
    \subfloat[Attacker's training dataset.]{
        \label{fig:a_train} 
        \includegraphics[width=3.2in]{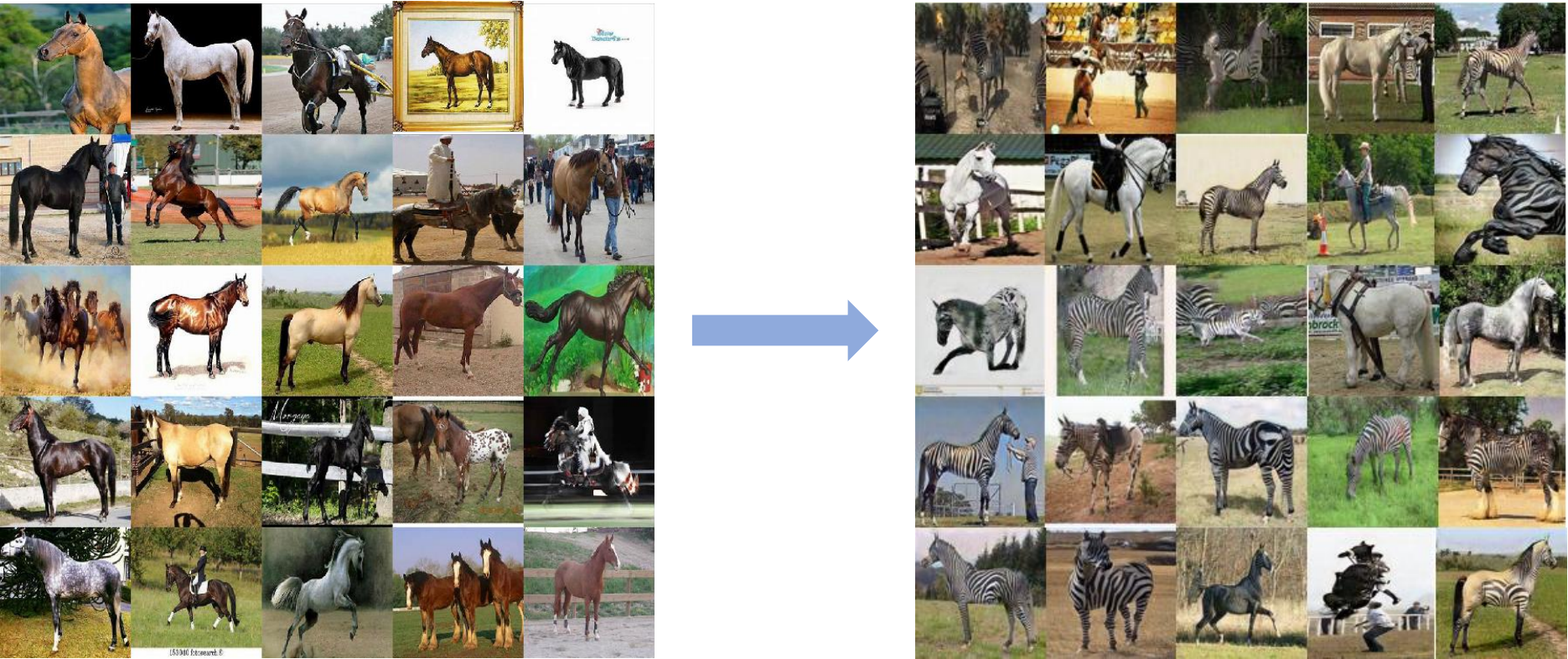}} 
    \caption{
    These two figures show a partial presentation of the training set for the victim model (Fig.~\ref{fig:v_train}) and the attack model (Fig.~\ref{fig:a_train}), respectively, 
    on the horse2zebra task. 
    }
    \label{fig:data_duibi} 
\end{figure}

\subsection{B.~~Visualization of the Domain Shift Problem}
\label{supp:domainshift}
In Fig.~\ref{fig:data_duibi}, we visualize the domain shift problem in the horse2zebra task. 
The source domains for both the training data of the victim model $D^X_V$ (upper left) and the attacker model $D^X_\mathcal{A}$ (lower left) consist of images of horses. However, there is a disparity between $D^X_V$ and $D^X_\mathcal{A}$. The victim's source domain  primarily consists of images of horses captured in the wild, whereas the attack dataset includes a more diverse set of horse images, such as sculptures and paintings. 
The domain shift problem leads to a situation where the targeted victim model is unable to accurately transfer certain horse images in the attack source domain $D^X_\mathcal{A}$ into zebra images (lower right). As a result, these 
low-quality images are incorporated into the attack model's training dataset, representing the target domain, and they introduce noise during the training process. This noise interferes with the final training outcome, making it harder for the attack model to successfully carry out the desired transformation from horse to zebra images. 

\begin{figure*}[t]
\begin{center}
  \includegraphics[width=6in]{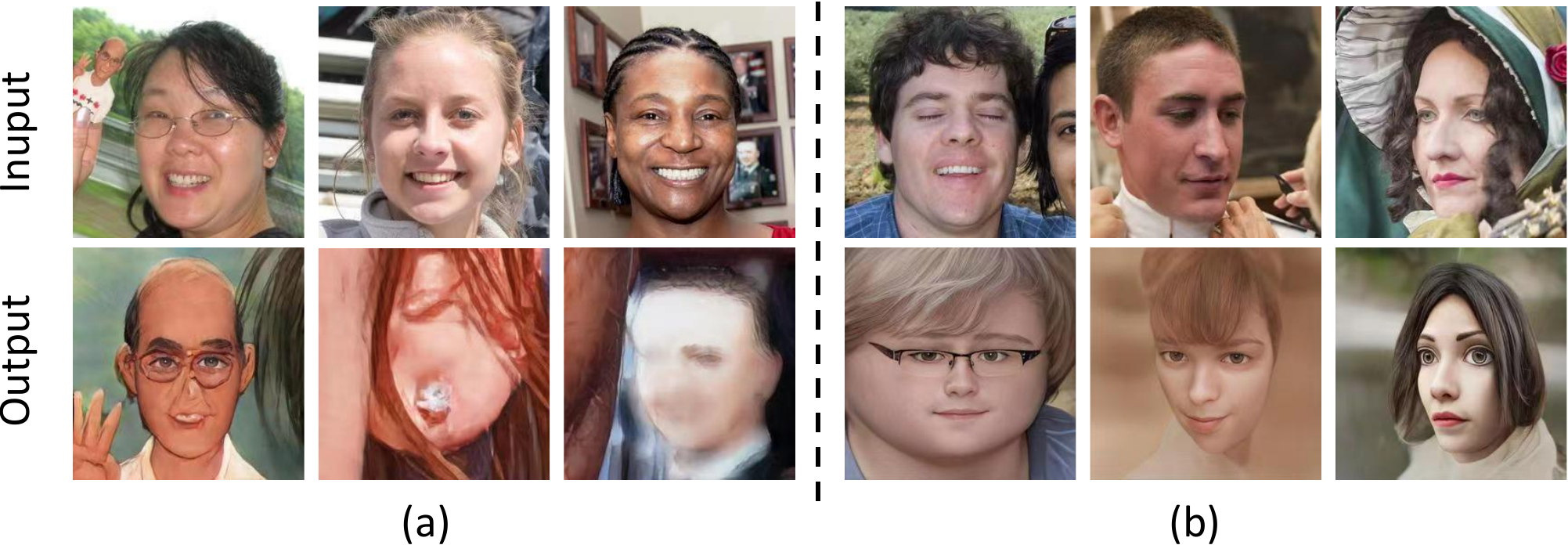}
  \caption{Example of the domain shift problem when querying the real-world commercial I2IT services. (a) face2cartoon (2d), (b) face2cartoon (3d).} 
  \label{fig:error}
\end {center}
 \end{figure*}

In Fig.~\ref{fig:error}, we show the domain shift problem when querying the real-world commercial I2IT services - Imglarger's human face cartoonization functions\footnotemark[1]. 

\begin{figure*}[t]
\begin{center}
  \includegraphics[width=6in]{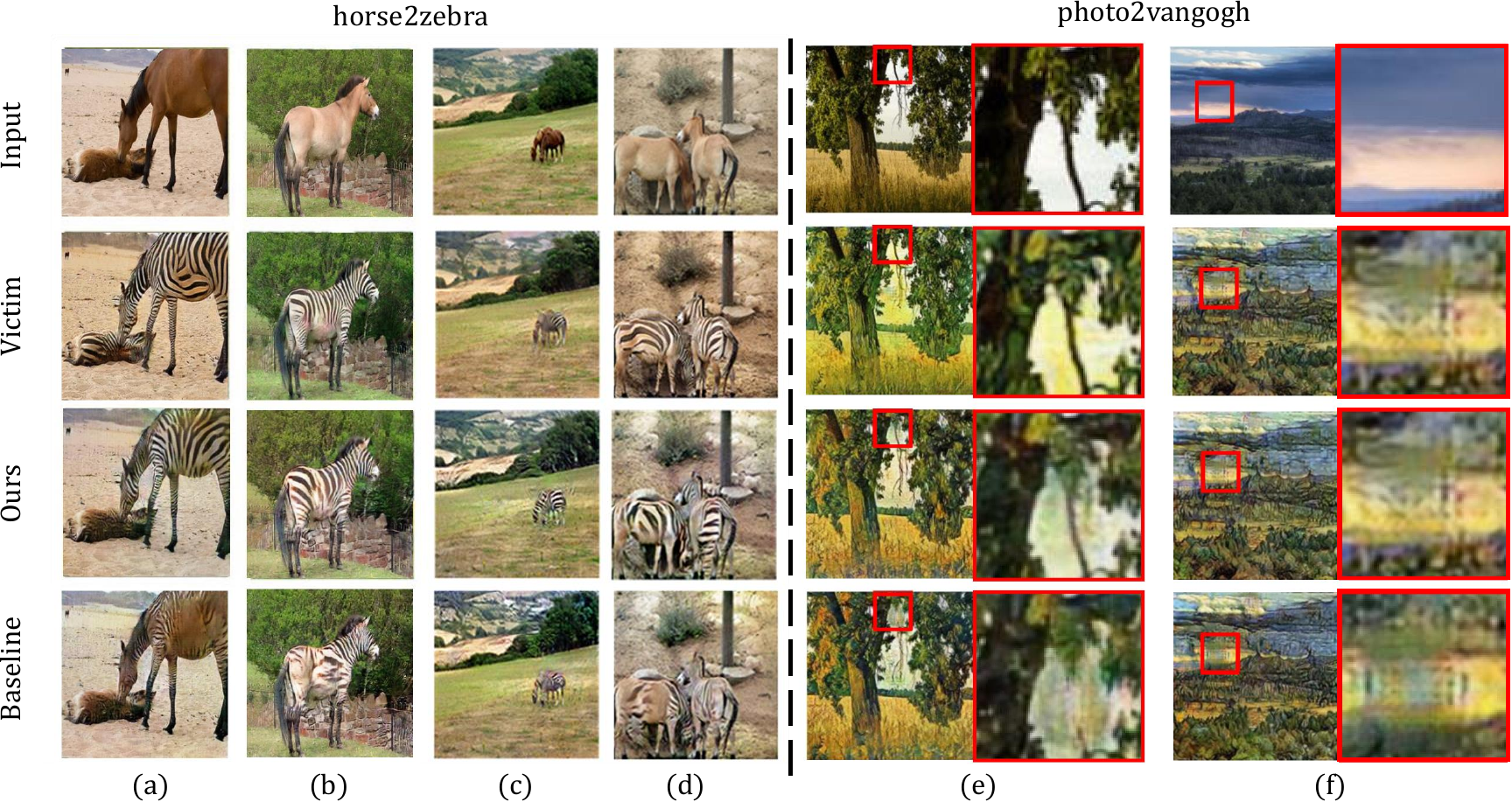}
  \caption{Qualitative results of our attack against the style transfer task. (a-d). horse2zebra, (e-f). photo2vangogh.} 
  \label{fig:duibi_supp}
\end {center}
 \end{figure*}

\begin{figure*}[!t]
\begin{center}
  \includegraphics[width=6in]{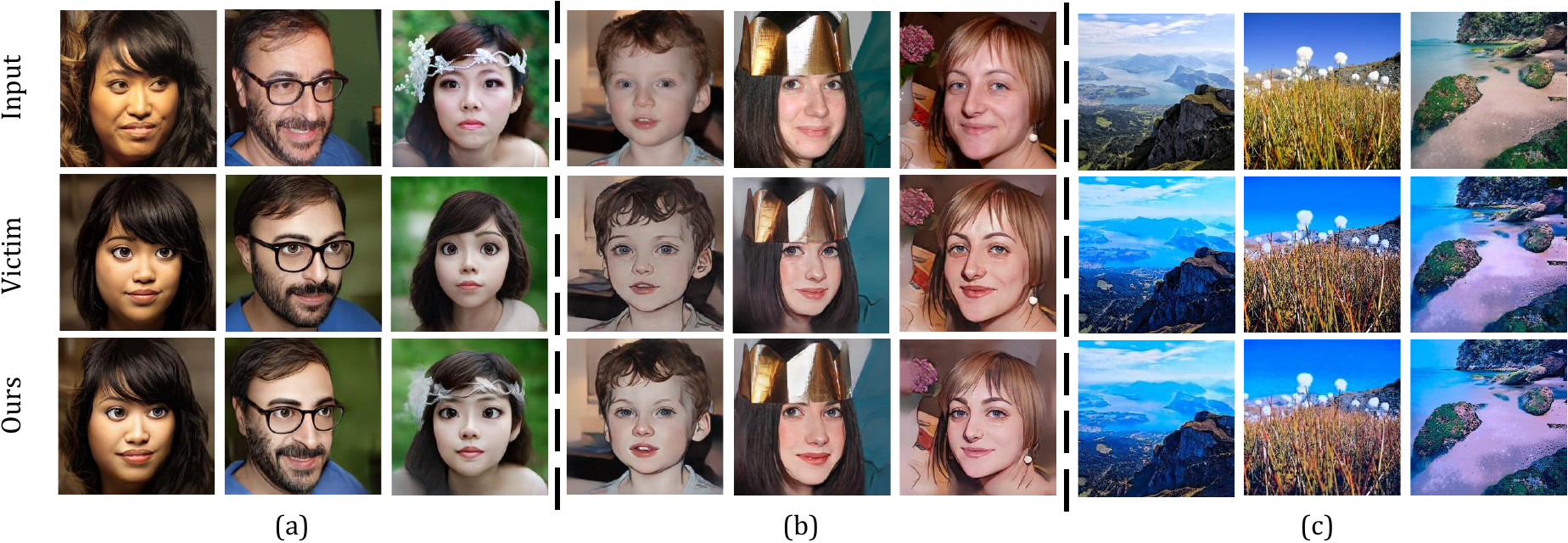}
  \caption{The performance of our attack against commercial
I2IT services. (a) Imglarger - face2cartoon 3d, (b) Imglarger - face2cartoon 2d, (c) Baidu AI Cloud - landscape2cartoon.} 
  \label{fig:real-duibi_supp}
\end {center}
 \end{figure*}

\subsection{C.~~Algorithmic supplementation}
\label{supp:algorithm}
In this section, we organize the specific process of training GAN using SAM into Algorithm~\ref{algorithm:SAM-GAN}. The weights of the generator and the discriminator are iteratively updated  (in line 3-7, and line 8-11, respectively). 

\begin{algorithm}[!t]
\caption{SAM-GAN} 
\label{algorithm:SAM-GAN}
\textbf{Input}: $G = \left\{G_1, \cdots, G_N \right\}$ and $D=\left\{D_1, \cdots, D_M\right\}$ are generators and discriminators in the backbone model, and their initial weights $\mathbf{w}^{G}=\left\{\mathbf{w}^{G_1}, \ldots,\mathbf{w}^{G_N}\right\}$, $\mathbf{w}^D=\left\{\mathbf{w}^{D_1}, \ldots,\mathbf{w}^{D_M}\right\}$, overall loss function is $L$, 
learning rate at time step $t$ is $\alpha_{t}$, loss sharpness neighborhood size $\rho_{G_i}>0$, $\rho_{D_j}>0$.\\
\textbf{Output}: $G$ and $D$ trained with SAM.\\
\textbf{Initialize}: $\mathbf{w}^G$ and $\mathbf{w}^D$ randomly
\begin{algorithmic}[1]
\WHILE{not converged}
\STATE Sample $b=\left\{(x^{(1)}, y^{(1)}), \cdots, (x^{(b)}, y^{(b)})\right\}$\\

\FOR {$G_i\in G$}
\STATE $\epsilon(\mathbf{w}^{G_i})=\rho_{G_i}\cdot\frac{\nabla_{\mathbf{w}^{G_i}}L(\mathbf{w})}{\|\nabla_{\mathbf{w}^{G_i}}L(\mathbf{w})\|_2}$\\
\STATE $g(\mathbf{w}^{G_i})=\nabla_{\mathbf{w}^{G_i}}L({\mathbf{w}^{G_i}})|_{\mathbf{w}^{G_i}+\epsilon(\mathbf{w}^{G_i})}$\\
\STATE $\mathbf{w}_{t+1}^{G_i}=\mathbf{w}_t^{G_i}-\alpha_{t}\cdot g(\mathbf{w}^{G_i})$ 
\ENDFOR

\FOR {$D_j\in D$}
\STATE $\epsilon(\mathbf{w}^{D_j})=\rho_{D_j}\cdot\frac{\nabla_{\mathbf{w}^{D_j}}L_j(\mathbf{w})}{\|\nabla_{\mathbf{w}^{D_j}}L_j(\mathbf{w})\|_2}$\\
\STATE $g(\mathbf{w}^{D_j})=\nabla_{\mathbf{w}^{D_j}}L_j({\mathbf{w}^{D_j}})|_{\mathbf{w}^{D_j}+\epsilon(\mathbf{w}^{D_j})}$\\
\STATE $\mathbf{w}_{t+1}^{D_j}=\mathbf{w}_t^{D_j}+\alpha_{t}\cdot g(\mathbf{w}^{D_j})$ 
\ENDFOR
\ENDWHILE
\STATE \textbf{return} $G, D$
\end{algorithmic}
\end{algorithm}

\subsection{D.~~Experiment Settings}
\label{supp:expsettings}
\subsubsection{Dataset Details}
We use the following three datasets as attack datasets:
\begin{itemize}
    \item Animal10~\cite{horse_datasets}. This dataset contains approximately 28,000 animal images. For our horse2zebra task, we only use 2,623 images from the "horse" folder as our attack dataset.
    \item Landscape~\cite{Landscape_datasets}. This dataset includes thousands of landscape photos collected from Flickr. When using this dataset, we randomly select 2,000 images as our attack dataset for the photo2vangogh task.
    \item Anime~\cite{Hayao_and_Shinkai_datasets}. In the super-resolution task, we utilize 256x256 images from works by Makoto Shinkai and Hayao Miyazaki. We randomly select 2,000 images from this dataset as our attack dataset for the super-resolution task. To simulate the blurring effects after image scaling and facilitate later comparison, we compress the images by a factor of four and then use bicubic interpolation to upscale them by four times.
\end{itemize}

\subsubsection{Implementation details}

In our experiments, we set batch size as 16. We augment the data using random clipping and random horizontal flipping. 
We use Adam as the optimizer for the baseline method while combining SAM and Adam as the optimizer for our method. 
In the style transfer task, the sharpness radius $\rho$ is set to 0.05, 
and $p$ is set to 2. In the super-resolution task, the sharpness radius $\rho$ is set to 0.03, 
and $p$ remains 2. 
For the wavelet regularization term $\alpha$, to keep it from overly interfering with the normal operation of the other loss components, we gradually increase it up to 0.15 and 0.25 in the style transfer and super-resolution task, respectively.

\subsection{E.~~More Experimental Results}
\label{sec:MoreResults}

\subsubsection{Additional Results for Comparing the Baseline and Our Attack with Different Number of Queries}
\begin{table}[t!]
\centering
\resizebox{0.47\textwidth}{!}{%
\begin{tabular}{| c | c | c | c |c|c | c |}
\hline
    \multirow{2}{*}{Task}& \multirow{2}{*}{Method} 
    & \multirow{2}{*}{\# Query} 
    &\multicolumn{2}{|c|}{$R_\text{capability}$} &\multicolumn{2}{|c|}{$R_\text{fidelity}$} \\
\cline{4-7}
     &  &  
     & FID$\downarrow$ & KID$\downarrow$ & FID$\downarrow$ & KID$\downarrow$ \\
\hline
        \multirow{3}{*}{\centering h2z\textsuperscript{$\dagger$}} &Victim   
        &NA  
        & 63.08 & 1.21±0.09 & NA & NA  \\
\cline{2-7}
        &Artist-Copy 
        &2623 
        & 115.09 & 5.59±0.39 & 75.65&1.93±0.25 \\
\cline{2-7}
        &Artist-Copy
        &5372 
        &   86.88 & 3.26±0.27 & 59.96&0.65±0.12 \\
\cline{2-7}
        &Ours 
        &2623 
        & \textbf{82.63} & \textbf{2.55±0.24} & \textbf{57.87}& \textbf{0.32±0.09}  \\
\hline
        \multirow{3}{*}{\centering p2v\textsuperscript{$\dagger$}} &Victim   
        &NA     
        & 109.78 & 2.73±0.43 &NA & NA \\
\cline{2-7}
        ~&Artist-Copy 
        &2000
        & 112.31 & 3.01±0.41 & 59.83 & 1.70±0.27\\
\cline{2-7}
        ~&Artist-Copy 
        &4000
        & 120.55 & 3.26±0.27 &64.27052 & 2.05±0.32\\
\cline{2-7}
        ~&Ours 
        &2000 
        & \textbf{110.00} & \textbf{2.66±0.37}& \textbf{59.09} &\textbf{1.45±0.25}  \\
\hline
\end{tabular}
}

\begin{flushleft}
\scriptsize
\textsuperscript{$\dagger$} We use h2z and p2v to represent horse2zebra and photo2vangogh. 
\end{flushleft}
\caption{\label{tabs1} Additional results for comparing the baseline and our attack with different number of queries against the style transfer tasks.} 
\end{table}

\begin{table}[t!]
\centering
\resizebox{0.47\textwidth}{!}{%
\begin{tabular}{| c | c | c |c|c | c |}
\hline
     \multirow{2}{*}{Method} 
    & \multirow{2}{*}{\# Query} 
    &\multicolumn{2}{|c|}{$R_\text{capability}$} &\multicolumn{2}{|c|}{$R_\text{fidelity}$} \\
\cline{3-6}
     &    
     & PSNR$\uparrow$ & LPIPS$\downarrow$ & PSNR$\uparrow$ & LPIPS$\downarrow$ \\
\hline
        Victim   
        &NA  
        & 34.56 & 0.067 & NA & NA  \\
\hline
        Artist-Copy 
        &2000 
        & 27.67  & 0.177 & 27.89& 0.156 \\
\hline
        Artist-Copy
        &4000 
        & 28.44 & 0.092 & 28.56  & 0.080\\
\hline
        Ours 
        &2000 
        & \textbf{29.56 } & \textbf{0.087} & \textbf{  29. 88 }& \textbf{0.077}  \\
\hline
\end{tabular}
}

\caption{\label{tab:super} Additional results for comparing the baseline and our attack with different number of queries against the super-resolution task.} 
\end{table}

\paragraph{Style Transfer} 
Our attack necessitates a significantly smaller number of queries (2-3k queries) compared to the baseline (with an average of 4-5k queries). 
As shown in Table~\ref{tabs1}, in the horse2zebra task, the {FID/KID} scores of our attack using 2,623 queries reach {82.63/2.55 for $R_\text{capability}$ and 57.87/0.32 for $R_\text{fidelity}$}, surpassing the Artist-Copy method which employs 5,372 queries. 
Likewise, in the photo2vangogh task, our attack using half the number of queries (\ie, 2,000 queries) achieving {110.00/2.66 and 59.09/1.45} FID/KID scores for $R_\text{capability}$ and $R_\text{fidelity}$, respectively. In contrast, the Artist-Copy method, {employing 4,000 queries, only attains FID/KID values of 120.55/3.26 for $R_\text{capability}$ and 64.27/2.05 for $R_\text{fidelity}$}. 

\paragraph{Super-resolution} 
Table~\ref{tab:super} shows the results of comparing the baseline and our attack with different number of queries against the super-resolution task. The PSNR/LPIPS scores of our attack using 2,000 queries achieve 29.56/0.087 and 29.88/0.077 for  $R_\text{capability}$ and $R_\text{fidelity}$, respectively. In comparison, the Artist-Copy requires 4,000 queries while only attaining PSNR/LPIPS values of 28.44/0.092 for  $R_\text{capability}$ and 28.56/0.080 for $R_\text{fidelity}$.

\subsubsection{Additional Qualitative Results of Our Attack Against the Style Transfer Task}
Fig.~\ref{fig:duibi_supp} shows more qualitative experimental results of our attack against the style transfer tasks.

\subsubsection{Additional Qualitative Results of Our Attack Against the Real-world Commercial I2IT Services}
Fig.~\ref{fig:real-duibi_supp} shows more experimental results of style transfer against the real-world commercial I2IT services\footnotemark[1].

\end{document}